\documentstyle[aps,prl,multicol,epsf]{revtex}

\def\lemma#1{\noindent{\bf Lemma$~#1$} \qquad}

\def\proof{\noindent{\bf Proof} \qquad}

\def\R{\hbox{\rm I \kern-5pt R}}

\def\Tr{{\rm{Tr}}}

\title{\rightline{ \small DAMTP-1999-76 \normalsize} \rightline{ 
\small quant-ph/9906006 \normalsize}
\rightline{} \centerline{Non-Contextual Hidden Variables and Physical Measurements} }
\author{ Adrian Kent}
\address{
Centre for Quantum Computation,
Clarendon Laboratory, Department of Physics, \\
University of Oxford, Parks Road, Oxford OX1 3PU, U.K. \\
\vskip 5pt
and 
\vskip 5pt
Department of Applied Mathematics and 
Theoretical Physics, University of Cambridge,\\ 
Silver Street, Cambridge CB3 9EW, U.K.${}^1$\\
} 

\date{1 June 1999}

\begin{document}
\maketitle
\begin{abstract}
For a hidden variable theory to be indistinguishable from quantum
theory for finite precision measurements, it is enough that its 
predictions agree for some measurement within the range of precision.
Meyer has recently pointed out that the Kochen-Specker theorem, which 
demonstrates the impossibility of a deterministic hidden variable
description of ideal spin measurements on a spin 1 particle, can thus
be effectively nullified if only finite precision measurements are
considered.  
We generalise this result: it is possible to ascribe consistent outcomes 
to a dense subset of the set of projection 
valued measurements, or to a dense subset of the set of positive
operator valued measurements, on any 
finite dimensional system.  
Hence no Kochen-Specker like contradiction can rule out hidden
variable theories indistinguishable from quantum theory by 
finite precision measurements in either class. 
\end{abstract}

\vskip10pt
${}^1$ Permanent address

\begin{multicols}{2}

\section{introduction}\label{introduction}

The experimental evidence against local hidden variable theories
is compelling and fundamental theoretical arguments weigh
heavily against those 
non-local hidden variable theories proposed to date,
but the question of hidden variables is still interesting
for at least two reasons. 

First, we want to distinguish strongly held theoretical beliefs
from established facts.  
Since there are reasonably sound 
(though clearly not universally persuasive) theoretical
arguments against {\it every} interpretation of quantum
theory proposed to date, we need to be 
particularly cautious about distinguishing belief from fact 
where quantum foundations are concerned. 
In particular, we still need to pin down precisely what types 
of hidden variable theories can or cannot be excluded by
particular theoretical arguments.  

Second, 
recent discoveries in quantum information theory and quantum computing
give new interest to some foundational questions.  
For example, we would like
to know in precisely what senses a quantum state carries
information and can be regarded as a computer, and how
it differs in these respects from classical analogues. 
From this perspective, as Meyer\cite{Meyer} has emphasized, 
questions about the viability of hidden
variable models for a particular process translate into 
questions about the classical simulability of some particular
aspect of quantum behaviour, and are interesting 
independently of the plausibility of the relevant models as 
physical theories.  

Either way, we need to distinguish arguments based on idealised
measurements, which can be specified precisely, from arguments
based on realistic physical measurements, which are always of 
finite precision.  Since we cannot precisely specify measurements, 
it is conceivable that all the measurements we can carry 
out actually belong to some subset of the full class of measurements
allowed by the quantum formalism.  That subset must be dense,
assuming that any finite degree of precision can be attained in 
principle, but it need not have positive measure.  Any no-go
theorem relying on a model of measurement thus has a potential
loophole which can be closed only if the theorem still holds
when measurements are restricted to a dense subset.  
From the point of view of quantum computation, the
precision attainable in a measurement is a computational 
resource: specifying infinite precision requires
infinite resources and prevents any useful comparison with
discrete classical computation.  

We consider here whether the predictions of quantum theory
can be replicated by a hidden variable model in which the 
outcomes of measurements are pre-determined by truth values
associated to the relevant operators.  
We first review the case of infinite precision with standard
von Neumann, i.e. projection valued, measurements. 
The question then is whether there is a consistent way 
of ascribing truth values $p(P) \in \{ 0, 1 \}$ to 
the projections in such a way as to determine a unique
outcome for any projection valued measurement. 
That is, is it possible to find a truth function $p$
such that if $\{ P_i  \}$ is a projective decomposition
of the identity then precisely one of the $P_i$ has truth
value $1$?  To put it formally, does there exist a truth
function $p$ such that 
\begin{equation}\label{one}
\sum_i p (P_i) = 1 
{\rm~if~} \sum_i P_i = I \, ? 
\end{equation} 

The Kochen-Specker (KS) theorem\cite{BellGleason,KochenSpecker,Specker} 
shows that the answer is no for systems 
whose Hilbert space has dimension greater than two.  
The general result follows from the result for projections in
three-dimensional real space, and so can be proved by
exhibiting finite sets of projections in $R^3$ for which a
truth function satisfying (\ref{one}) is demonstrably impossible.
Kochen and Specker gave the first example\cite{KochenSpecker} of 
such a set, and
some simpler examples were later found by Peres.\cite{Peres,Peresbook}
An independent proof was given by Bell,\cite{BellGleason} 
who noted that, by an argument of Gleason's,\cite{Gleason} 
(\ref{one}) implies a minimal finite separation
between projectors with truth value $1$ and $0$, which is 
impossible since both values must be attained. 

We could ask the same question about positive operator valued
measurements.  That is, does there exist a truth
function $p$ on the positive operators such that 
\begin{equation}\label{two}
\sum_i p (A_i) = 1 
{\rm~if~} \sum_i A_i = I \, ? 
\end{equation} 
Obviously, since projection valued measurements are special cases of
positive operator valued measurements, the KS theorem 
still applies.

Returning to the case of projections, we want to know whether
the restriction to finite precision could make a difference.
Our hypothesis, recall, is that a finite precision measurement 
could correspond to a measurement of some particular 
projective decomposition in the precision range, a 
decomposition whose projections do indeed have hidden
pre-assigned truth values.
The relevant question then is: is there a physically sensible
truth function $p$ defined on a dense subset $S_1$
of the space of all projections such that (\ref{one}) holds on a 
dense subset $PD_1$ of the space of all projective decompositions?  
This requires in particular that the set $S_2$ of projections belonging to 
decompositions in $PD_1$ must satisfy $S_2 \subseteq S_1$.
By physically sensible, we mean that the subsets
$S_3$ and $S_4$ of projections $P$ in $S_2$ for which $p(P)=1$
and $p(P) = 0$ respectively are both dense in the space of
all projections, so as to avoid the possibility of 
contradiction by experiments of sufficiently high precision.  

We first consider the case treated in the proof of the KS theorem, 
one-dimensional projections on $R^3$.   The possibility of 
hidden variable models evading the KS theorem was first
considered by Pitowsky.\cite{Pitowsky}
Meyer\cite{Meyer} has given a
very pretty example of 
a truth function $p$ defined on the subset $S^2 \cap Q^3$
of projections defined by rational vectors, which satisfies
(\ref{one}) for all orthogonal triples.  
Meyer's elegant proofs,\cite{Meyer} using earlier work of
Godsil and Zaks,\cite{GodsilZaks} show 
that all the necessary denseness
conditions hold and hence that the KS proof is indeed
nullified if we restrict attention to finite precision 
measurements.

Meyer's result shows that the KS theorem cannot be directly 
applied in the finite precision case.  However, it does not imply that
the theorem itself is false, or that no similar no-go theorem
can be found.  Even in
the case of three dimensional systems, this requires
an example of a truth function that
satisfies (\ref{one}) for a dense subset of the triads
of projections on $C^3$ rather than $R^3$.  A more complete
argument requires an example of a physically sensible 
truth function satisfying
(\ref{one}) for a dense subset of the projective decompositions of
the identity on $C^n$.  More generally still, since all physical
measurements are actually positive operator valued, a complete
defence against KS-like arguments requires an example of
a physically sensible truth function satisfying 
(\ref{two}) for a dense subset of the positive operator decompositions of
the identity on $C^n$.

We give such examples here.  First, some notation.
Define the one-dimensional 
projection $P_{r_1 , \ldots , r_{2n}}$ on $C^n$ to be the projection 
onto the vector $N ( r_1 + i r_2 , \ldots , r_{2n-1} + i r_{2n} )$,
where the $r_i$ are real and not all zero and the
normalisation constant 
obeys $N^{-2} = \sum_{i=1}^{2n} r_i^2 $.
Call $P_{r_1 , \ldots , r_{2n}}$ {\it true} if all the
$r_i$ are rational and non-zero and if, writing $r_i = p_i / q_i$ 
we have that $q_1$ is divisible by $3$ and none
of the other $q_i$ are.  Here, and throughout, 
any fractions we write are taken to be in lowest terms.
Call an n-tuple $\{ Q_1 , \ldots , Q_n \}$ of orthogonal
one-dimensional projections {\it suitable} if at least one of the 
$Q_i$ is true.  
If $P$ belongs to a suitable n-tuple but is not true, 
call $P$ {\it false}.  (Note that a projection need not be
either true or false.)
Define $p (P) =1$ if $P$ is true and $p(P) =0$ if $P$ is false.
\vskip10pt
\lemma1 A suitable n-tuple contains precisely one true projection.

\proof  If $P$ and $Q$ are both true projections, the corresponding
vectors have inner product of the form $ (a/9) + (p/q) + i (r/s)$, 
where $3$ is not a factor of $q$.  The real part thus cannot vanish,
so $P$ and $Q$ cannot be orthogonal.

\vskip10pt
\lemma2 The true projections are dense in the space of all
one-dimensional projections.

\proof  Given any one-dimensional 
projection $P_{r_1 , \ldots , r_{2n}}$ we can find an arbitrarily
close approximation $P_{r'_1 , \ldots , r'_{2n}}$ with rational
$r'_i = p'_i / q'_i$.  If $q'_1$ is not divisible by $3$, we can find an
arbitrarily close rational approximation $r''_1 = p''_1 / q''_1$ to $r'_1$ 
with $q''_1$ divisible by $3$, for example by taking 
$p''_1 = 3 N p'_1 + 1$ and $q''_1 = 3 N q'_1$ for a
sufficiently large integer $N$.  
Similarly, if any of the $q'_i$ for $i>1$ are divisible
by $3$, we can find arbitrarily close rational approximations
$r''_i = p''_i / q''_i$ to $r'_i$ with $q''_i$ not 
divisible by $3$, for example by taking 
$p''_1 = N p'_1 $ and $q''_1 =  N q'_1 +1$ for 
a sufficiently large integer $N$.

\vskip10pt
\lemma3 The suitable n-tuples are dense in the space of all n-tuples
of orthogonal projections.

\proof Given any n-tuple $\{ P_1 , \ldots , P_n \}$, 
choose one of the projections,
say $P_1$.  As above, we can find an arbitrarily
close approximation to $P_1$ by a true projection $Q$.
Let $U$ be a rotation in $SU(n)$ which rotates $P_1$ to $Q$ 
such that $ | U - I | = ( \Tr ( (U - I) ( U^{\dagger} - I ) ) )^{1/2} $ 
attains the minimal value for 
such rotations.  The compactness of $SU(n)$ ensures that such
a $U$ exists, though it need not be unique, and the minimal
value tends to zero as $Q$ tends towards $P$.  
The projections $\{ U P_1 , \ldots , U P_n \}$ form
a suitable n-tuple, and this construction gives n-tuples of
this type arbitrarily close to the original.

\vskip10pt
\lemma4  The false projections are dense in the space of
all one-dimensional projections.

\proof Given any projection $P$, choose an n-tuple to which it
belongs, and let $Q$ be another projection in that n-tuple.
By the construction above, we can find arbitrarily close
n-tuples in which the projections approximating $Q$ are
true.  The projections approximating $P$ are thus
false.

This concludes the argument for measurements defined by n-tuples of
one-dimensional projections.  For completeness, though, we also
consider degenerate von Neumann measurements, corresponding to 
decompositions of
the identity into general orthogonal projections.  
The construction above generalises quite simply.
Fixing the basis as before, we can write each projection as a 
matrix: $P = N ( a_{ij} + i b_{ij} )_{i,j = 1}^n$, where the
$a_{ij}$ and $b_{ij}$ are real and $N$
is some normalisation constant. 
Consistently with our earlier definitions for one-dimensional
projections, we can define $P$ to be true if it can be 
written in this form with all the $a_{ij}$
and $b_{ij}$ rational and non-zero and if $a_{11}$ is then the
only one which, when written in lowest terms, has denominator
divisible by $9$.  Clearly if 
$P$ and $Q$ are both true then $\Tr (PQ) \neq 0$, so they cannot
be orthogonal.  We can thus define suitable projective decompositions
and false projections as above, and all the earlier arguments
run through with trivial modifications. 

At this stage a comment on measurement theory is required.  
The KS theorem assumes the traditional von Neumann definition 
of measurement, in which measurement projects the quantum state 
onto an eigenspace of the relevant observable.  In more
realistic modern treatments, a measurement causes an action
on the quantum state by positive operators, which may
but need not be close to projections.  One could, indeed,
realistically base measurement theory only on positive operator 
valued measurements in which the positive operators are not
projections, for example stipulating that all positive
operators involved must be of maximal rank.  If so, the 
original KS theorem becomes irrelevant, though it can easily
be modified to deal with these cases.  It seems more natural,
though, to either allow any precisely specified positive
operator decomposition, whether or not it includes projections,
or else to consider general finite precision positive operator
valued measurements.  If all precisely specified positive
operators are included, then of course the KS theorem applies.
On the other hand, as we now show, the finite precision loophole
also exists for positive operator measurements. 

We need new definitions for positive operators.
Again fixing a basis, we can write a positive operator 
as a matrix:
$A  = ( a_{ij} + i b_{ij} )_{i,j = 1}^n$, where the $a_{ij}$
and $b_{ij}$ are real, so that $a_{ij} = a_{ji}$ and 
$b_{ij} = - b_{ji}$.  We say that $A$ is {\it true} if
$a_{11} = r_1 + r_2 \sqrt{2}$, with $r_1$ and $r_2$ both
rational and $r_2$ positive, and that a projective decomposition
$I = \sum_i A_i$ of the identity into positive operators is
{\it suitable} if precisely one of the $A_i$ is true.  
$A$ is {\it false} if it belongs to a suitable decomposition
but is not true.  Under this definition, every $A$ is either true
or false.
Define the truth function $p$ by
setting $p (A) =1$ if $A$ is true and $p(A) =0$ if $A$ is false.
Clearly $p$ satisfies (\ref{two}) on suitable decompositions.
Clearly, too, true and false operators are dense in the
space of positive operators, and suitable decompositions are dense
in the space of all positive operator decompositions.  
Hence the desired result holds.

Note that these last definitions, restricted to projections, give
another example of a physically sensible truth function
satisfying (\ref{one}).  
The two different constructions perhaps help to illustrate the 
large scope for examples of this sort.
There is nothing particularly special about either our 
constructions or those of
Ref. \cite{Meyer}: the possibility of
closing the finite precision loophole by any KS-type argument can be  
refuted in many different ways.

It follows from the above examples that 
non-contextual hidden variable theories cannot be excluded 
by theoretical arguments of the KS type once the imprecision in
real world experiments is taken into account.  
This does not, of course, imply that such theories
are very plausible, or that the particular
constructions we give are capable of producing a
physically interesting hidden variable theory.
Nor does the discussion affect the situation regarding
local hidden variable theories, which 
can be refuted by experiment, modulo reasonable 
assumptions.\cite{BellEPR,CHSH,aspectetal}

\vskip10pt
\noindent{\bf Acknowledgments}
I am grateful to Philippe Eberhard for suggesting clarifications
in the presentation and to the Royal Society for financial support.

\end{multicols}

\end{document}